# Indicatori comuni del PNRR e *framework* SDGs: una proposta di indicatore composito*

Fabio Bacchini[1,2], Lorenzo Di Biagio[2],
Giampiero M. Gallo[1,3], Vincenzo Spinelli[2†]

**Abstract**

The main component of the NextGeneration EU (NGEU) program is the Recovery and Resilience Facility (RRF), spanning an implementation period between 2021 and 2026. The RRF also includes a monitoring system: every six months, each country is required to send an update on the progress of the plan against 14 common indicators, measured on specific quantitative scales.

The aim of this paper is to present the first empirical evidence on this system, while, at the same time, emphasizing the potential of its integration with the sustainable development framework (SDGs). Our proposal is to develop a first linkage between the 14 common indicators and the SDGs which allows us to produce a composite index (SDGs-RRF) for France, Germany, Italy and Spain for the period 2014-2021. Over this time span, widespread improvements in the composite index across the four countries led to a partial reduction of the divergence.

The proposed approach represents a first step towards a wider use of the SDGs for the assessment of the RRF, in line with their use in the European Semester documents prepared by the European Commission.

In this respect, Italy's experience is valuable, given the inclusion of well-being and sustainability indicators in public finance assessments and the availability of the NRRP-SDGs dashboard prepared by Istat and the State General Accounting Department (RGS).

**Sommario**

La principale componente del programma NextGeneration EU (NGEU) è il Recovery and Resilience Facility (RRF), che ha una durata dal 2021 al 2026. L'RRF comprende anche un sistema di monitoraggio: semestralmente, ciascun paese è tenuto a inviare un aggiornamento sugli avanzamenti del piano rispetto a 14 indicatori comuni, misurati su specifiche scale quantitative.

Obiettivo del presente lavoro è di illustrare tale sistema, presentando da un lato le prime evidenze empiriche, ma sottolineando, dall'altro, le potenzialità delle integrazioni con il *framework* dello sviluppo sostenibile (SDGs). La proposta da noi sviluppata, incentrata su un primo raccordo tra i 14 indicatori comuni e gli SDGs, ha permesso di elaborare un indice composito (SDGs-RRF) per Francia, Germania, Italia e Spagna per il periodo 2014-2021. Nel periodo considerato, i diffusi miglioramenti dell'indice composito tra i quattro paesi hanno portato a una riduzione parziale della divergenza.

L'approccio proposto rappresenta un primo step verso un più ampio utilizzo degli SDGs per la valutazione dell'RRF, in linea con il loro utilizzo all'interno dei documenti sul semestre europeo predisposti dalla Commissione europea.

In quest'ottica, l'esperienza italiana è di particolare valore, in considerazione dell'inserimento degli indicatori di benessere e sostenibilità all'interno dei documenti di finanza pubblica e della disponibilità del cruscotto PNRR-SDGs predisposto da Istat e Ragioneria Generale dello Stato.

*Parole chiave:* PNRR, valutazione delle politiche, benessere e sostenibilità, indicatori compositi.

# 1 Introduzione

La principale componente del programma NGEU è il Dispositivo per la Ripresa e Resilienza (Recovery and Resilience Facility, RRF), che ha una durata di sei anni, dal 2021 al 2026, e una dimensione totale di 672,5 miliardi di euro (312,5 sovvenzioni, i restanti 360 miliardi prestiti a tassi agevolati).

Nella predisposizione dell'RRF, particolare attenzione è stata dedicata alla valutazione degli effetti[1] che le misure contenute in ciascun piano nazionale potranno avere sull'evoluzione economica, sociale e ambientale di ciascun paese, illustrandone i progressi nel tempo rispetto a ciascuno dei sei pilastri delle aree di intervento[2]. Per l'Italia, com'è noto, l'RRF ha assunto il nome di Piano Nazionale per la Ripresa e la Resilienza (PNRR).

Ai fini della valutazione, la Commissione europea ha predisposto un apposito Scoreboard per fornire *an overview of how the implementation of the Recovery and Resilience Facility (RRF) and the national recovery and resilience plans is progressing*[3]. Lo Scoreboard presenta una ampia articolazione dedicata al monitoraggio del raggiungimento delle *milestone* e dei *target* concordati con ciascun paese[4].

Lo scoreboard include inoltre un set di 14 *indicatori comuni* per tutti i paesi, allo scopo di monitorare il progresso nella realizzazione degli obiettivi previsti dall'RRF, il cui dettaglio è discusso nel paragrafo successivo.

La scelta di un set di indicatori comuni appare condivisibile rispetto alla loro facilità di lettura e trasparenza e anche ai criteri di individuazione che, improntati al monitoraggio dei 6 pilastri, trovano anche un ampio riscontro all'interno del più ampio dibattito sul *beyond GDP* [Stiglitz et al., 2009]. In quest'ottica è opportuno ricordare come l'approccio *beyond GDP* ha portato al disegno e realizzazione di *framework* internazionali dedicati al benessere [si veda, ad esempio, Bacchini et al., 2021, con riferimento all'Italia] e allo sviluppo sostenibile (Sustainable Development Goals – SDGs), coordinato a livello internazionale dalle Nazioni Unite[5].

Tuttavia, la comune ispirazione che ha legato il processo di selezione dei 14 indicatori al dibattito sullo sviluppo sostenibile non esprime ancora un quadro organico di relazioni con il sistema degli indicatori SDGs che costituiscono, invece, una componente consolidata all'interno dei documenti elaborati dalla Commissione europea (Commissione Europea [2022])[6]. Anche tenendo conto di successivi aggiustamenti alle declinazioni specifiche degli RRF nazionali, si tratta di avere un supporto alla valutazione di politiche, vale a dire una strumentazione adatta ad analizzare il contributo degli interventi finanziati al raggiungimento degli obiettivi prefissati. Da questo punto di vista, l'introduzione di SDGs nel quadro RRF aiuterebbe nell'interpretare l'impatto dei piani nazionali sui fenomeni ritenuti rilevanti.

---

[1] Si veda l'art. 30 del Regolamento (UE) 2021/241 del Parlamento Europeo e del Consiglio del 12 febbraio 2021 (d'ora in poi, Regolamento, che istituisce il Dispositivo per la ripresa e la resilienza - DRR).

[2] 1 - transizione verde; 2 - trasformazione digitale; 3 - crescita intelligente, sostenibile e inclusiva, che comprenda coesione economica, occupazione, produttività, competitività, ricerca, sviluppo e innovazione, e un mercato interno ben funzionante con PMI forti; 4 - coesione sociale e territoriale; 5 - salute e resilienza economica, sociale e istituzionale, al fine, fra l'altro, di rafforzare la capacità di risposta alle crisi e la preparazione alle crisi; 6 - politiche per la prossima generazione, l'infanzia e i giovani, come l'istruzione e le competenze.

[3] Si veda il sito della Commissione europea.

[4] Le milestones rappresentano obiettivi qualitativi mentre i target si riferiscono a obiettivi quantitativi.

[5] Con riferimento alle Nazioni Unite, si rimanda all'apposito sito, mentre per l'Italia al portale dedicato al tema realizzato dall'Istat.

Obiettivo di questo lavoro è di suggerire una prima ipotesi di raccordo, sviluppando una mappatura sia per una singola Missione–Componente del PNRR che per l'insieme dei 14 indicatori comuni. L'armonizzazione proposta consente di sfruttare l'insieme degli indicatori SDGs, permettendo confronti in serie storica e tra paesi, ampliando il perimetro dei risultati riportati sullo scoreboard.

La mappatura proposta ha consentito di elaborare un indice composito per i 14 indicatori, denominato SDGs-RRF, ottenuto utilizzando le rispettive proxies all'interno degli SDGs. L'indice è calcolato per Francia, Germania, Italia e Spagna per il periodo 2014-2021. I risultati presentati forniscono anche una misura dell'evoluzione della convergenza tra i paesi considerati, rafforzando le evidenze disponibili in letteratura (si vedano, ad esempio, Campos and Macchiarelli [2021], Bacchini et al. [2020b]).

L'approccio proposto rappresenta un primo passo verso un più ampio utilizzo degli SDGs per la valutazione degli RRF, in linea con l'inserimento degli indicatori di benessere e sostenibilità in Italia all'interno dei documenti di finanza pubblica e la disponibilità del dashboard PNRR-SDGs[7] predisposto da Istat e Ragioneria Generale dello Stato (cfr. anche Istat 2022, par. 1.4), iniziative che pongono l'Italia come paese guida per questi sviluppi.

Il lavoro è organizzato come segue. Nella Sezione 2 si presenta il quadro generale riferito agli indicatori europei, analizzando la loro declinazione all'interno delle articolazioni del PNRR in termini di intensità e diffusione, anche rispetto alle amministrazioni coinvolte. La Sezione 3 illustra, con un esempio, il funzionamento del sistema dei 14 indicatori, proponendo dapprima un'ipotesi di raccordo tra gli indicatori individuati e il sistema SDGs per singola componente e, successivamente, una mappatura dell'insieme dei 14 indicatori rispetto agli SDGs. In base alle corrispondenze individuate, nella Sezione 4 si elaborano gli indici compositi SDGs-RRF per i 4 paesi. Infine, nella Sezione 5 si prospettano alcune linee di sviluppo delle attività volte a rafforzare l'utilizzo del sistema SDGs per il monitoraggio del PNRR/RRF.

---

[6] In quest'ottica si veda anche il contributo fornito da Asvis [2023].
[7] Si veda il cruscotto Istat-RGS con gli indicatori di sostenibilità.

# 2 Il *framework* degli indicatori europei

## 2.1 Quadro generale

Il quadro di valutazione predisposto per il monitoraggio del PNRR/RRF è rappresentato dal cosiddetto obiettivo specifico[8], che consiste nella valutazione del raggiungimento formale dei traguardi e degli obiettivi collegati, nel caso italiano, a ciascuna delle 285 misure di investimento e di riforma.

Se l'obiettivo specifico è perseguito in stretta e trasparente cooperazione con gli Stati interessati, attraverso il sistema delle *milestone* e dei target, quello generale richiede uno sguardo di insieme più armonico rappresentato dal sistema degli indicatori comuni e dagli elementi dettagliati del quadro di valutazione della ripresa e della resilienza[9]. Gli indicatori comuni vengono compilati e diffusi due volte l'anno in accordo con il calendario del semestre europeo, attraverso uno specifico dashboard[10]. La descrizione dei 14 indicatori e la loro articolazione con i pilastri del RRF sono riportati nella Figura 1 insieme all'unità di misura utilizzata.

La scelta degli indicatori produce una diversa messa a fuoco degli effetti del dispositivo, privilegiando, verosimilmente, gli obiettivi la cui rendicontazione presenta una minore complessità. Del resto, come sottolineato più volte nella documentazione ufficiale, la misurazione degli indicatori identificati rappresenta il progresso conseguito dal singolo paese e non una distanza da uno specifico target.

Formalmente, ciascuno dei 14 indicatori può essere ulteriormente articolato in sotto classificazioni. Ad esempio, per la capacità operativa supplementare per energia rinnovabile si distingue tra la produzione di energia rinnovabile e quella di idrogeno, mentre per le infrastrutture per i combustibili alternativi si osservano i punti di ricarica e i punti di rifornimento (e, per questi, quelli per l'idrogeno sono separati).

Ulteriori disaggregazioni degli indicatori riguardano, ove possibile, le età (quattro classi 0-17, 18-29, 30-54, 55 e oltre) e il genere (M/F), mentre per le imprese l'articolazione è sviluppata lungo tre classi dimensionali (grande, media e piccola/micro) e, nello specifico, separatamente per sviluppo o adozione di tecnologie digitali. Il quadro complessivo include dunque 48 differenti misurazioni derivate dai 14 indicatori comuni.

In Italia, l'attribuzione degli indicatori a ciascuna delle 285 misure nelle quali è articolato il PNRR è stata realizzata attraverso il dialogo con le Amministrazioni titolari, nell'ambito del Tavolo di monitoraggio[11].

---

[8]si veda art. 4 del Regolamento.

[9]Si veda il Regolamento Delegato (UE) 2021/2106 della Commissione del 28 settembre 2021.

[10]La scadenza per la fornitura degli aggiornamenti da parte dei paesi è fissata al 28 febbraio e al 31 agosto.

[11]Si veda la circolare n. 34 del 17 ottobre 2022 dell'Unità di Missione NG-EU della RGS che specifica come gli indicatori comuni dovranno essere valorizzati a livello di singolo CUP. Il soggetto attuatore è responsabile della corretta alimentazione dei relativi dati, ferma restando la responsabilità e validazione della bontà delle informazioni in capo

Figura 1: Descrizione degli indicatori e allocazione nei pilastri

| | INDICATORI | PILASTRI | | | | | | Unità |
|---|---|---|---|---|---|---|---|---|
| | | 1 | 2 | 3 | 4 | 5 | 6 | |
| 1 | Risparmi sul consumo annuo di energia primaria | ■ | | ■ | | | | MWh/anno |
| 2 | Capacità operativa supplementare installata per l'energia rinnovabile | ■ | | ■ | | | | MW |
| 3 | Infrastrutture per i combustibili alternativi (punti di ricarica/rifornimento) | ■ | | ■ | | | | Punti |
| 4 | Popolazione che beneficia di misure di protezione contro inondazioni, incendi boschivi e altre catastrofi naturali connesse al clima | ■ | | | ■ | | | Abitanti |
| 5 | Abitazioni aggiuntive con accesso a Internet fornito attraverso reti ad altissima capacità | | ■ | | ■ | | | Abitazioni |
| 6 | Imprese beneficiarie di un sostegno per sviluppare o adottare prodotti, servizi e processi applicativi digitali | | ■ | ■ | | | | Imprese |
| 7 | Utenti di servizi, prodotti e processi digitali pubblici nuovi e aggiornati | | ■ | | | ■ | | Utenti/anno |
| 8 | Ricercatori che lavorano in centri di ricerca beneficiari di un sostegno | | | ■ | | | | Equivalenti a T/P |
| 9 | Imprese beneficiarie di un sostegno (tra cui piccole imprese, comprese le microimprese, medie e grandi imprese) | | | ■ | | | | Imprese |
| 10 | Numero di partecipanti in un percorso di istruzione o di formazione | | ■ | | ■ | | ■ | Persone |
| 11 | Numero di persone che hanno un lavoro o che cercano un lavoro | | | ■ | | | | Persone |
| 12 | Capacità delle strutture di assistenza sanitaria nuove o modernizzate | | | | ■ | ■ | | Persone/anno |
| 13 | Capacità delle classi nelle strutture per la cura dell'infanzia e nelle strutture scolastiche nuove o modernizzate | | | | ■ | | ■ | Persone |
| 14 | Numero di giovani di età compresa tra i 15 e i 29 anni che ricevono sostegno | | | | | | ■ | Persone |

Fonte: Elab. Corte dei conti sull'Allegato al Regolamento Delegato (UE) 2021/2106 della Commissione del 28/09/2021. I pilastri sono denominati come segue: 1 - transizione verde; 2 - trasformazione digitale; 3 - crescita intelligente, sostenibile e inclusiva, che comprenda coesione economica, occupazione, produttività, competitività, ricerca, sviluppo e innovazione, e un mercato interno ben funzionante con PMI forti; 4 - coesione sociale e territoriale; 5 - salute e resilienza economica, sociale e istituzionale, al fine, fra l'altro, di rafforzare la capacità di risposta alle crisi e la preparazione alle crisi; 6 - politiche per la prossima generazione, l'infanzia e i giovani, come l'istruzione e le competenze.

Ricordando che le 285 misure sono ripartite tra 220 investimenti (I) e 65 riforme (R), nella Tabella 1 si riporta il loro numero per ciascuna Missione/Componente (colonne "Misure"); sotto le colonne "di cui con IC", si dà conto delle misure associate ad almeno un indicatore comune. Il grado di coinvolgimento di almeno un indicatore per Missione/Componente è variabile, passando da casi di assenza di un indicatore di riferimento (M3C1), a tutte le misure di una M/C per le quali è presente almeno un indicatore (es. M2C3).

---

all'Amministrazione centrale titolare della misura.

Tabella 1: Misure PNRR per Missione e Componente e quelle a cui è associato almeno un indicatore comune.

| Comp | Descrizione Componente | Misure | | di cui con IC | | Con IC/ Tot. misure |
|---|---|---|---|---|---|---|
| | | I | R | I | R | |
| M1C1 | Digitalizzazione, innovazione e sicurezza nella PA | 30 | 20 | 20 | | 67% |
| M1C2 | Digitalizzazione, innovazione e competitività nel sistema produttivo | 18 | 2 | 12 | | 67% |
| M1C3 | Turismo e Cultura 4.0 | 36 | 2 | 28 | | 78% |
| M2C1 | Agricoltura sostenibile ed Economia Circolare | 8 | 3 | 6 | | 75% |
| M2C2 | Energia rinnovabile, idrogeno, rete e mobilità sostenibile | 24 | 5 | 18 | | 75% |
| M2C3 | Efficienza energetica e riqualificazione degli edifici | 4 | 1 | 4 | | 100% |
| M2C4 | Tutela del territorio e della risorsa idrica | 15 | 4 | 7 | | 47% |
| M3C1 | Investimenti sulla rete ferroviaria e sulla sicurezza stradale | 14 | 4 | | | 0% |
| M3C2 | Intermodalità e logistica integrata | 6 | 6 | 6 | | 100% |
| M4C1 | Potenziamento dell'offerta dei servizi di istruzione: dagli asili nido alle università | 13 | 10 | 13 | 2 | 100% |
| M4C2 | Dalla ricerca all'impresa | 11 | 1 | 11 | | 100% |
| M5C1 | Politiche per il lavoro | 5 | 2 | 5 | 1 | 100% |
| M5C2 | Infrastrutture sociali, famiglie, comunità e terzo settore | 13 | 2 | 9 | | 69% |
| M5C3 | Interventi speciali per la coesione territoriale | 8 | 1 | 2 | | 25% |
| M6C1 | Reti di prossimità, strutture e telemedicina per l'assistenza sanitaria territoriale | 5 | 1 | 2 | | 40% |
| M6C2 | Innovazione, ricerca e digitalizzazione del servizio sanitario | 10 | 1 | 9 | | 90% |
| | Totale per tipologia | 220 | 65 | 152 | 3 | 69% |
| | Totale | 285 | | 155 | | 54% |

Fonte: Elab. Corte dei conti sull'Allegato al Regolamento Delegato (UE) 2021/2106 della Commissione del 28/09/2021. Le colonne "Misure" riportano i totali delle misure per Missione e Componente distinte per Investimenti e Riforme; quelle "di cui con IC", i totali delle misure con almeno un indicatore associato. L'ultima colonna riporta la loro quota percentuale sul totale.

La mancata associazione di un intervento a un indicatore è legata alla sua natura e non alla sua importanza finanziaria: ad esempio, nell'ambito della Componente M2C4 (Tutela del territorio e della risorsa idrica), le misure M2C4I0401 (Investimenti in infrastrutture idriche primarie per la sicurezza dell'approvvigionamento idrico), M2C4I0402 (Riduzione delle perdite nelle reti di distribuzione dell'acqua, compresa la digitalizzazione e il monitoraggio delle reti), M2C4I0403 (Investimenti nella resilienza dell'agro-sistema irriguo per un migliore gestione delle risorse idriche) e M2C4I0404 (Investimenti in fognatura e depurazione) non sono associate ad alcuno dei quattordici indicatori anche se, nel complesso, lo stanziamento loro attribuito dal PNRR è pari a 4,38 miliardi di euro. Pur nella semplicità dell'esempio proposto, è evidente come la mancata associazione di un intervento agli indicatori comuni non offusca la sua rilevanza in termini di contributo ai fini del raggiungimento degli obiettivi generali di resilienza e di adattamento e contrasto al cambiamento climatico.

Una rappresentazione complementare si ottiene ricostruendo la distribuzione delle misure e delle 155 occorrenze (su 285) di almeno un indicatore comune tra le Amministrazioni titolari (Tabella 2): in termini di attenzione, per così dire, da dedicare agli indicatori comuni, 13 Amministrazioni sono responsabili di investimenti completamente associati ad almeno un indicatore, mentre il Ministero delle Infrastrutture e della Mobilità Sostenibili (MIMS, con 12 su 35 (il 34%), il Ministero dell'innovazione e transizione digitale (MITD, con 16 su 28, il 57%) e il Ministero della transizione ecologica (MITE, con 18 su 28, il 64%) registrano valori più contenuti del rapporto tra indicatori e misure. Queste Amministrazioni[12] svolgono un ruolo rilevante rispetto sia al valore dei finanziamenti affidati che all'impatto previsto per la sostenibilità, la transizione digitale e il cambiamento climatico.

Tabella 2: Distribuzione delle misure e degli indicatori per Amministrazione

| Ammin. Titolare | Misure I | Misure R | di cui con IC I | di cui con IC R | con IC Tot. misure |
|---|---|---|---|---|---|
| Giust. Amministrativa | 1 | | 1 | | 100% |
| MAECI - Min Aff Est | 1 | | 1 | | 100% |
| MEF - Min Eco e Fin | 1 | 6 | 1 | | 100% |
| MG - MIN GIUSTIZIA | 2 | 5 | 2 | | 100% |
| MI - Min Istruzione | 10 | 6 | 10 | 1 | 100% |
| MIC - Min Cultura | 23 | | 21 | | 91% |
| MIMS - Min Infr Mob Sost | 35 | 11 | 12 | | 34% |
| MINT - Min Interno | 5 | | 5 | | 100% |
| MIPAAF - Min Pol Agr | 4 | | 3 | | 75% |
| MISE - Min Svil Eco | 16 | 1 | 16 | | 100% |
| MITD - Min Inn Trans Dig | 28 | 4 | 16 | | 57% |
| MITE - Min Trans Eco | 28 | 12 | 18 | | 64% |
| MiTur - Min Turismo | 13 | 1 | 7 | | 54% |
| MLPS - Min Lav Pol S | 9 | 3 | 5 | 1 | 56% |
| MS - Min Salute | 15 | 2 | 11 | | 73% |
| MUR - Min Uni e Ric | 11 | 5 | 11 | 1 | 100% |
| PCM - Dip Prot Civil | 1 | | 1 | | 100% |
| PCM - Dip Sport | 1 | | 1 | | 100% |
| PCM - Min Aff Reg | 1 | | 1 | | 100% |
| PCM - Min Pari Opp | 1 | | 1 | | 100% |
| PCM - Min Pol Giov | 1 | | 1 | | 100% |
| PCM - Min Pub Amm | 9 | 5 | 5 | | 56% |
| PCM - Min Sud | 4 | 1 | 2 | | 50% |
| PCM-DISAB | | 1 | | | |
| PCM-SEGR_GEN | | 2 | | | |
| Totale per tipo | 220 | 65 | 152 | 3 | 69% |
| Totale | 285 | | 155 | | 54% |

Fonte: Elab. Corte dei conti sull'Allegato al Regolamento Delegato (UE) 2021/2106 della Commissione del 28/09/2021. Le colonne "Misure" riportano i totali delle misure per Missione e Componente distinte per Investimenti e Riforme; quelle "di cui con IC", i totali delle misure con almeno un indicatore associato. L'ultima colonna riporta la loro quota percentuale sul totale.

La relazione tra misure e indicatori comuni può essere ulteriormente dettagliata, osservando il numero delle volte che ciascun indicatore, ovvero una sua declinazione, viene considerato per ogni missione/componente. Le frequenze osservate (Tabella 3) possono essere interpretate come numero di valori che complessivamente confluiranno nel quadro di valutazione previsto dal Regolamento per ciascuna missione e componente del PNRR: alcuni di questi valori saranno ripetuti quando l'indicatore comune è associato a più misure della stessa componente.

---

[12]Per molti la denominazione ufficiale è stata modificata con il cambio di legislatura a fine 2022.

La mappa delle associazioni permette di caratterizzare l'impatto atteso del PNRR in termini di monitoraggio. L'indicatore C10, partecipazione a percorsi di istruzione o formazione, declinato per genere e per età, rappresenta l'indicatore più diffuso (288 sui 646 casi di associazione), con riferimento alle componenti che prevedono espressamente un innalzamento del livello di preparazione in campi ritenuti strategici, particolarmente per quanto riguarda le competenze digitali.

L'indicatore C9, numero di imprese che ricevono sostegno, articolato in classi dimensionali, è particolarmente diffuso: anche in questo caso digitalizzazione e innovazione, turismo, energia rinnovabile e mobilità sostenibile, e collegamento tra ricerca e impresa costituiscono le componenti più coinvolte. L'indicatore C14, giovani che ricevono supporto, articolato per genere, assume una discreta rilevanza tra le componenti ed è utilizzato per il monitoraggio degli interventi di digitalizzazione e innovazione per la PA e per il servizio sanitario, e quelli relativi all'istruzione.

Alcuni indicatori assumono una rilevanza limitata tra le componenti, come ad esempio C5, abitazioni raggiunte da internet ad alta capacità, associato unicamente 3 volte in M1C2 (Digitalizzazione, innovazione e competitività nel sistema produttivo), C12 presente 2 volte in ciascuna nelle due componenti della Missione M6 -Salute e C13 (capacità delle classi in edifici scolastici e per l'infanzia) che compare a proposito della riqualificazione energetica degli edifici (M2C3) e del potenziamento dei servizi di istruzione (M4C1).

Tabella 3: Associazione tra Missioni, Componenti e indicatori comuni

| Comp | C1 | C2 | C3 | C4 | C5 | C6 | C7 | C8 | C9 | C10 | C11 | C12 | C13 | C14 | Totale |
|---|---|---|---|---|---|---|---|---|---|---|---|---|---|---|---|
| M1C1 | | | | | | | 13 | | | 88 | | | | 16 | 117 |
| M1C2 | | | | | 3 | 3 | 1 | | 27 | | | | | | 34 |
| M1C3 | 9 | | | | | | 15 | | 33 | 16 | | | | 4 | 77 |
| M2C1 | 2 | 3 | | | | 6 | | | 9 | 8 | | | | 2 | 30 |
| M2C2 | 3 | 6 | 8 | | | | | | 33 | | | | | | 50 |
| M2C3 | 4 | | | | | | | | | | | | 1 | | 5 |
| M2C4 | 1 | 1 | | 5 | | | 2 | | | | | | | | 9 |
| M3C2 | 1 | | 2 | | | | 5 | | | | | | | | 8 |
| M4C1 | 2 | | | | | | 2 | 2 | | 56 | | | 3 | 16 | 81 |
| M4C2 | | | | | | 12 | | 18 | 24 | | | | | 4 | 58 |
| M5C1 | 1 | | | | | | 1 | | 6 | 24 | 8 | | | 6 | 46 |
| M5C2 | 6 | | | | | | 1 | | 3 | 24 | | | | 4 | 38 |
| M5C3 | | | | | | | | | 3 | 8 | | | | 2 | 13 |
| M6C1 | | | | | | | | | | | | 2 | | | 2 |
| M6C2 | | | | | | | 2 | 2 | | 64 | | 2 | | 8 | 78 |
| Totale | 29 | 10 | 10 | 5 | 3 | 21 | 42 | 22 | 138 | 288 | 8 | 4 | 4 | 62 | 646 |

Fonte: Elab. Corte dei conti sull'Allegato al Regolamento Delegato (UE) 2021/2106 della Commissione del 28/09/2021.

La distribuzione degli indicatori può essere osservata anche per Amministrazione titolare (Tabelle 4). Quattro Amministrazioni titolari responsabili sono chiamate a rendicontare un solo indicatore su un'unica misura associata: il MEF (su C9 – imprese beneficiarie di un sostegno) e i Dipartimenti della Protezione Civile, quello dello Sport e quello delle Politiche Giovanili della PCM, (rispettivamente, il C4 – popolazione che beneficia di misure di protezione contro gli eventi connessi ai cambiamenti climatici; il C1 – risparmi energetici e C14 - giovani che ricevono supporto). Le altre Amministrazioni titolari forniscono una rendicontazione più articolata fornendo informazioni su più indicatori. Ad esempio, il MITE è chiamato ad aggiornare sette indicatori comuni, mentre altri quattro ministeri (Istruzione, Cultura, Interno, Lavoro e PS) ne compileranno cinque ciascuno.

Tabella 4: Associazione tra Amministrazioni e indicatori comuni

| Amministrazione titolare | C1 | C2 | C3 | C4 | C5 | C6 | C7 | C8 | C9 | C10 | C11 | C12 | C13 | C14 | Totale |
|---|---|---|---|---|---|---|---|---|---|---|---|---|---|---|---|
| Giust Amministrativa | | | | | | | | | | 8 | | | | 2 | 10 |
| MAECI - Min Aff Est | | | | | | 3 | | | 3 | | | | | | 6 |
| MEF - Min Eco e Fin | | | | | | | | | 3 | | | | | | 3 |
| MG - MIN GIUSTIZIA | 1 | | | | | | | | | 16 | | | | 2 | 19 |
| MI - Min Istruzione | 3 | | | | | | 2 | | | 48 | | | 4 | 6 | 63 |
| MIC - Min Cultura | 3 | | | | | | | 14 | | 15 | 16 | | | 4 | 52 |
| MIMS - Min Infr Mob | 4 | | 6 | | | | 5 | | 3 | | | | | | 18 |
| MINT - Min Interno | 4 | 1 | 1 | 1 | | | | | 3 | | | | | | 10 |
| MIPAAF - Min Pol Agr | | 1 | | | | 6 | | | 9 | | | | | | 16 |
| MISE - Min Svil Eco | | | | | | | 12 | 1 | 4 | 48 | | | | | 65 |
| MITD - MIN INNOV TEC | | | | | 3 | | 11 | | | 32 | | | | 8 | 54 |
| MITE - Min Trans Eco | 4 | 7 | 3 | 3 | | | 2 | | 18 | 8 | | | | 2 | 47 |
| MiTur - Min Turismo | 6 | | | | | | 1 | | 18 | | | | | | 25 |
| MLPS - Min Lav Pol S | 2 | | | | | | 2 | | | 48 | 8 | | | 8 | 68 |
| MS - Min Salute | | | | | | | 2 | 2 | | 64 | | 4 | | 8 | 80 |
| MUR - Min Uni e Ric | | | | | | | | 16 | 12 | 8 | | | | 14 | 50 |
| PCM - Dip Prot Civil | | | | 1 | | | | | | | | | | | 1 |
| PCM - Dip Sport | 1 | | | | | | | | | | | | | | 1 |
| PCM - Min Aff Reg | 1 | 1 | | | | | | | | | | | | | 2 |
| PCM - Min Pari Opp | | | | | | | | | 3 | | | | | | 3 |
| PCM - Min Pol Giov | | | | | | | | | | | | | | 2 | 2 |
| PCM - Min Pub Amm | | | | | | | 2 | | | 32 | | | | 4 | 38 |
| PCM - Min Sud | | | | | | | | | 3 | 8 | | | | 2 | 13 |
| Totale | 29 | 10 | 10 | 5 | 3 | 21 | 42 | 22 | 138 | 288 | 8 | 4 | 4 | 62 | 646 |

Fonte: Elab. Corte dei conti sull'Allegato al Regolamento Delegato (UE) 2021/2106 della Commissione del 28/09/2021.

# 3 Dagli indicatori comuni europei agli SDGs

La definizione di un sistema di monitoraggio armonizzato a livello europeo rappresenta un importante elemento per comprendere i risultati raggiunti in relazione al finanziamento ricevuto. Tuttavia, un'integrazione tra il nuovo sistema e quelli già disponibili come ad esempio il *framework* SDGs, potrebbe rappresentare un'ulteriore opportunità anche ai fini della valutazione dell'impatto dei fondi a valere sui piani nazionali RRF.

## 3.1 La lettura degli indicatori europei

Il disegno del sistema di monitoraggio europeo prevede uno specifico Scoreboard che fornisce *an overview of how the implementation of the Recovery and Resilience Facility (RRF) and the national recovery and resilience plans is progressing*[13].

---
[13] Le informazioni sullo Scoreboard sono disponibili selezionando questo link.

Il sito dedicato allo Scoreboard contiene una sezione dedicata agli indicatori comuni nella quale si riportano i risultati per ciascun indicatore in tutti i paesi dell'Unione europea (Figura 2). I valori sono cumulati rispetto al tempo, cioè si riferiscono al periodo che va dall'inizio dell'attività dell'RRF all'ultimo trimestre disponibile.

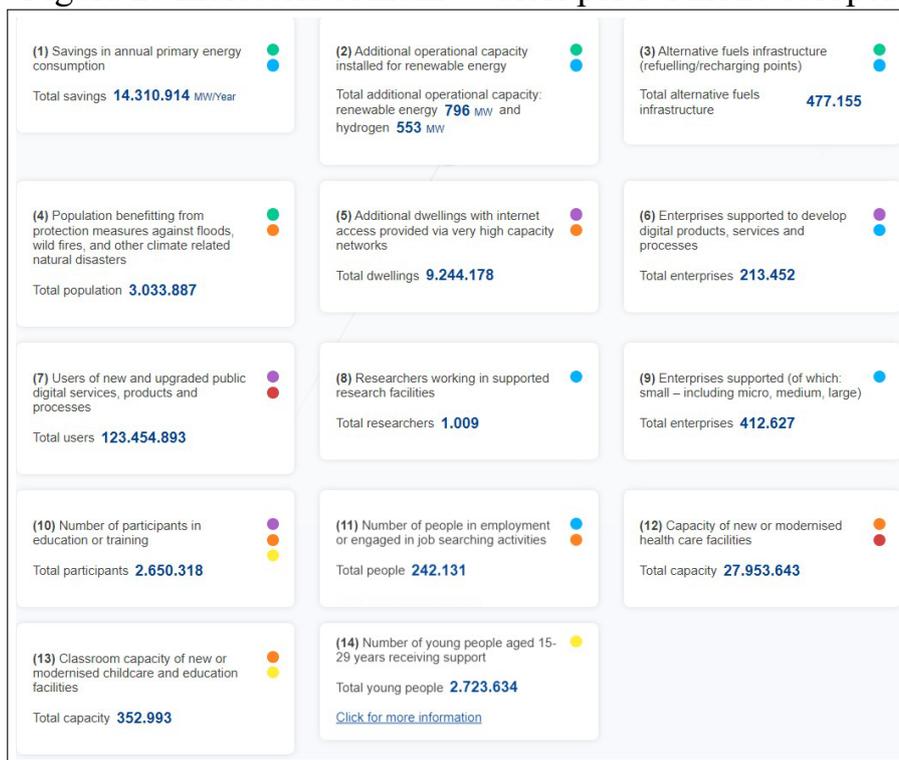

Figura 2: Indicatori comuni - valori per l'Unione europea

Fonte: Commissione europea - Recovery and resilience Scoreboard

Considerando come esempio l'indicatore C7, utenti di servizi, prodotti e processi digitali pubblici nuovi e aggiornati, il valore riportato è pari a 123,5 milioni di utenti. Selezionando il link disponibile per l'indicatore è possibile osservare la ripartizione temporale: 37,9 milioni di utenti sono stati contabilizzati fino a dicembre 2021 mentre gli altri 85,6 milioni successivamente.

La disaggregazione per paese, riportata come percentuale sul totale, presenta una forte eterogeneità. Considerando il valore dell'indicatore per l'ultimo periodo, denominato I semestre 2022, la percentuale degli utenti è particolarmente elevata per la Spagna (53%) e per la Francia (51,9%), assumendo valori marcatamente superiori sia al dato dell'Italia (14,6%) che a quello della Germania (0,2%). E' opportuno notare che la somma delle quote può essere superiore al 100% poiché, come riportato nella legenda, *the same person can use the service multiple times, in which case they would be counted multiple times*.

Rispetto al quadro europeo, i dati disponibili sul sito di Italiadomani forniscono il dettaglio della situazione per l'Italia con riferimento al II semestre del 2022.

Tabella 5: Indicatori comuni Italia - II semestre 2022

| Codice | Frequenza | Unità di misura |
|---|---|---|
| C5 | 88.412 | Abitazioni |
| C6 | 2.274 | Imprese |
| C7 | 12.059.648 | Utenti/anno |
| C8 | 120.193 | Equivalenti a T/P |
| C9 | 94.772 | Imprese |
| C10 | 164.761 | Persone |
| C10 - digitale | 71.261 | Persone |
| C11 | 2.448 | Persone |
| C14 | 149.786 | Persone |

Fonte: Italiadomani

La prima osservazione riguarda il numero di indicatori per i quali è disponibile una quantificazione che costituisce un sottoinsieme dei 14 disponibili. Nel II semestre 2022 sono stati riportati informazioni su 8 indicatori dei 14 previsti mentre non ci sono informazioni per gli indicatori riferiti all'energia (C1, C2 e C3), protezione del suolo (C4), sanità (C12) e infanzia (C13).

L'elenco delle Amministrazioni che hanno fornito dati nel secondo semestre include la Giustizia amministrativa (Consiglio di Stato e Tar), il MAE, il MIUR, il Ministero del lavoro e politiche sociali, il Ministero del turismo, il Ministero dell'istruzione e del merito, il Ministero della giustizia, il Ministero delle imprese e del made in Italy e la Presidenza del consiglio (dipartimento politiche giovanili, coesione, trasformazione digitale e funzione pubblica).

Il quadro delineato evidenzia da un lato la difficoltà di interpretare gli avanzamenti dei progetti del PNRR in termini dei 14 indicatori, sia a livello internazionale che nazionale. Da un lato, emerge la necessità di motivare l'elevata eterogeneità presente nelle analisi per il singolo indicatore (si veda l'esempio per l'indicatore C7). Dall'altro risulta ancora difficile realizzare un'articolata analisi dei movimenti degli indicatori mancando una consolidata esperienza di utilizzo degli stessi nelle analisi.

## 3.2 La dimensione della sostenibilità

Come visto precedentemente nell'esempio sul numero di utenti di servizi, prodotti e processi digitali pubblici nuovi, è possibile conoscere l'ammontare totale del valore dell'indicatore per semestre, la sua composizione per paese e, verosimilmente, l'ammontare dei fondi RRF utilizzati per incrementare tale numero. All'interno di questo scenario è, quindi, possibile effettuare, ad esempio, dei confronti in termini di costo del finanziamento per singolo utente nei diversi paesi. Questa prospettiva offre quindi un'alternativa rispetto ai tradizionali sistemi di monitoraggio basati su indicatori nei quali stabilire la relazione con gli effetti della misura politica sotto osservazione appare più difficile.

Tuttavia, il vantaggio di costruire una relazione diretta tra finanziamento ed evoluzione dell'indicatore in termini di flusso, così come proposto dallo schema europeo sui 14 indicatori, presenta al momento almeno tre difficoltà. La prima criticità consiste nell'introdurre un ulteriore sistema che appare, almeno al momento, sconnesso dai *framework* di misurazione già disponibili: alcuni indicatori sembrano peraltro riconducibili a fenomeni già oggetto di analisi statistica a livello europeo, ma rilevati annualmente. In quest'ottica, la scelta di basarsi su indicatori di flusso semestrale potrebbe costituire un elemento di criticità interpretativa.

In secondo luogo, il sistema degli indicatori comuni fornisce una specifica disaggregazione sia per classi di età e per genere ma non richiede, almeno al momento, una disaggregazione geografica regionale. la mera declinazione nazionale risulta limitativa in presenza di obiettivi che mirano alla riduzione dei divari territoriali, come nel caso del PNRR. Tuttavia, nel caso italiano, i dati a disposizione per i singoli interventi (CUP) permetterebbero una ricostruzione più dettagliata a livello spaziale.

Infine, il Dispositivo trova motivazione nell'attenuazione dell'impatto economico e sociale della pandemia, migliorando la sostenibilità e la resilienza delle società e favorendo la transizione green e digitale, elementi questi, che hanno forti punti di contatto con il *framework* dello sviluppo sostenibile (SDGs)[14]. Da questo punto di vista, i piani nazionali e il sistema degli indicatori comuni appaiono ancora non integrati con questo framework, che costituisce, invece, uno dei tratti caratterizzanti la documentazione del semestre europeo e, in particolare, dei Country report elaborati dalla Commissione europea. Il mancato ancoraggio del sistema dei 14 indicatori al

---

[14] Come è noto il sistema SDGs è articolato in 17 goals e, nel Rapporto 2023 per l'Italia, contiene 372 misure statistiche, di cui 342 uniche, cioè associate ad un unico goal: Goal 1 – Porre fine a ogni forma di povertà nel mondo; Goal 2 – Porre fine alla fame, raggiungere la sicurezza alimentare, migliorare la nutrizione e promuovere un'agricoltura sostenibile; Goal 3 – Assicurare la salute e il benessere per tutti e per tutte le età; Goal 4 – Istruzione di qualità per tutti - fornire un'educazione di qualità, equa ed inclusiva e promuovere opportunità di apprendimento continuo per tutti; Goal 5 – Raggiungere l'uguaglianza di genere l'empowerment di tutte le donne e le ragazze; Goal 6 – Garantire a tutti la disponibilità e la gestione sostenibile dell'acqua e delle strutture igienico sanitarie; Goal 7 – Assicurare a tutti l'accesso a sistemi di energia economici, affidabili, sostenibili e moderni; Goal 8 – Promuovere una crescita economica duratura, inclusiva e sostenibile, un'occupazione piena e produttiva e un lavoro dignitoso per tutti; Goal 9 – Costruire una infrastruttura resiliente e promuovere l'innovazione e una industrializzazione equa, responsabile e sostenibile; Goal 10 – Ridurre le disuguaglianze fra i paesi e al loro interno; Goal 11 – Rendere le città e gli insediamenti umani inclusivi, sicuri, resilienti e sostenibili; Goal 12 – Garantire modelli sostenibili di produzione e di consumo; Goal 13 – Adottare misure urgenti per combattere il cambiamento climatico e le sue conseguenze; Goal 14 – Conservare e utilizzare in modo sostenibile gli oceani, i mari e le risorse marine per uno sviluppo sostenibile; Goal 15 – Proteggere, ripristinare e favorire un uso sostenibile degli ecosistemi terrestri, gestire in modo sostenibile le foreste, combattere la desertificazione, arrestare e invertire il degrado del territorio e arrestare la perdita di biodiversità; Goal 16 – Promuovere società pacifiche e inclusive per uno sviluppo sostenibile; rendere disponibile l'accesso alla giustizia per tutti e creare organismi efficaci, responsabili e inclusivi a tutti i livelli; Goal 17 – Rafforzare i mezzi di attuazione e rinnovare il partenariato mondiale per lo sviluppo sostenibile.

*framework* SDG è stato sottolineato anche dalla Commissione europea[15], in uno studio che ha analizzato i piani nazionali di ripresa e resilienza (NRRP), evidenziando come in pochi casi sia presente un chiaro collegamento con il *framework* SDG, ovvero un esplicito riferimento ai relativi indicatori.

Ad esempio, considerando l'Italia, il Country report 2022 riporta una disamina dei progressi raggiunti dal paese rispetto ai 17 Goal (Figura 3). Il giudizio complessivo indica un diffuso miglioramento della posizione italiana, sottolineando comunque alcune criticità[16]. Nello stesso documento si sottolinea come le misure contenute nel PNRR riferite alle politiche attive sul mercato del lavoro e alla formazione, alla coesione sociale e territoriale, ai servizi sociali e all'inclusione sono attese avere un impatto significativo sulla performance italiana verso gli SDGs.

Figura 3: Progressi verso gli SDGs in Italia negli ultimi 5 anni

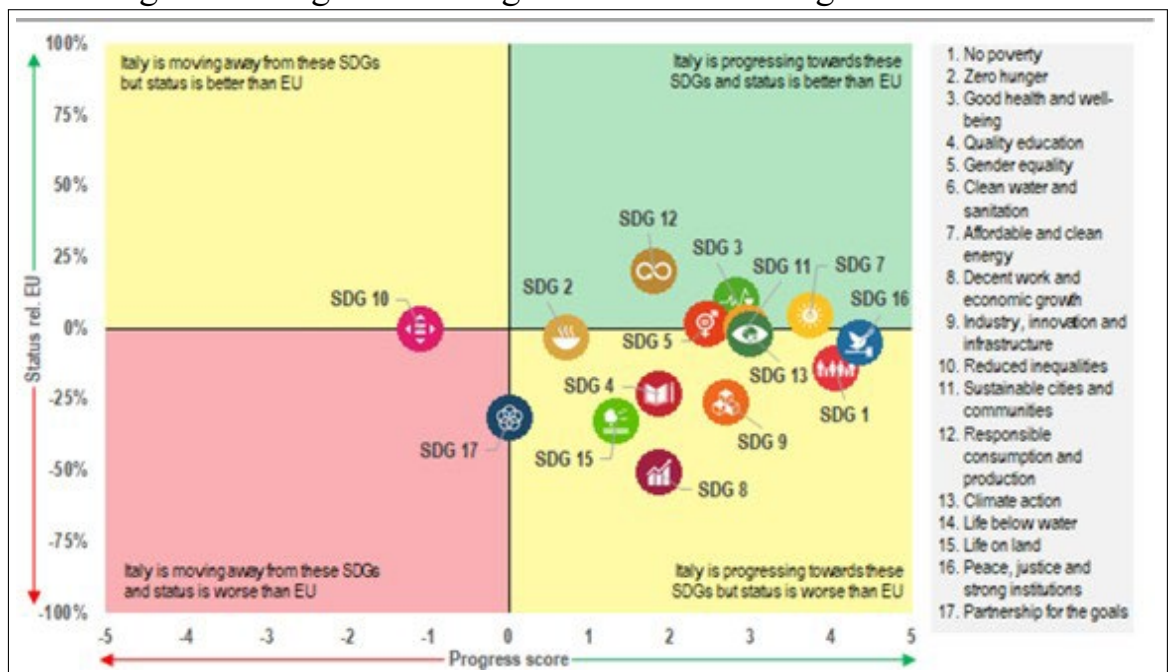

Fonte: Commissione europea - 2022 Country Report - Italy

Nell'ottica di stimolare una riflessione più ampia a livello europeo, rivolta a ricucire questa distanza, l'esperienza italiana potrebbe rappresentare una *best practice*. A settembre 2022, infatti, la Ragioneria generale dello Stato ha diffuso insieme all'Istituto nazionale di statistica (Istat) il *Monitoraggio delle misure del PNRR attraverso gli indicatori di sviluppo sostenibile (SDGs) e dell'Agenda 2030*[17].

---

[15]European Committee of the regions, 2022, Synergies between the Sustainable Development Goals and the National Recovery and Resilience Plans – Best Practices from Local and Regional Authorities. Nel summary si riporta che "The majority of Member States merely mention the SDGs implicitly with very few Member States explicitly and clearly linking NRRP components to the SDGs" … "the use of SDG indicators in the NRRPs is limited to very few Member States, and for only a few SDGs, limiting both comparability between the plans as well as effective monitoring of their contribution to progress towards the SDGs".

[16]Nel testo si rileva come in Italia siano aumentate le disuguaglianze (Goal 10) e come le quota di persone povere o in stato di grave deprivazione sia ancora più elevata della media europea (Goal 1). Anche con riferimento ai NEET, al gap di genere sui salari (Goal 5 e 8) e al livello di educazione (Goal 4) si evidenziano decise difficoltà.

Per ogni misura, componente e sub-misura è quindi possibile osservare contemporaneamente sia il flusso semestrale di uno dei 14 indicatori proposti dalla Commissione, sia gli indicatori SDGs associati[18]. Ad esempio, per la componente M5C1, denominata *politiche del lavoro*, la proposta di monitoraggio della Commissione europea, nell'articolazione concordata tra le amministrazioni pubbliche italiane è associata agli indicatori comuni presentati nella Tavola 6, mentre quelli estratti dal sistema SDG sono presentati nella Tavola 7 unitamente al goal di riferimento.

Tabella 6: Lista indicatori comuni afferenti alla componente M5C1 – politiche del lavoro

| Indicatori comuni | |
|---|---|
| Codice | Descrizione |
| C1 | Risparmi sul consumo annuo di energia primaria |
| C7 | Utenti di servizi, prodotti e processi digitali pubblici nuovi e aggiornati |
| C9 | Imprese beneficiarie di un sostegno (tra cui piccole imprese, comprese le microimprese, medie e grandi imprese) |
| C10 | Numero di partecipanti in un percorso di istruzione o di formazione |
| C11 | Numero di persone che hanno un lavoro o che cercano un lavoro |
| C14 | Numero di giovani di eta` compresa tra i 15 e i 29 anni che ricevono sostegno |

Fonte: elaborazioni su dati Italiadomani

Tabella 7: Indicatori SDGs afferenti alla componente M5C1 – politiche del lavoro

| Descrizione indicatore SDGs | Goal di riferimento |
|---|---|
| Uscita precoce dal sistema di istruzione e formazione | 4 |
| Giovani che non lavorano e non studiano (NEET) | 8 |
| Partecipazione alla formazione continua | 4 |
| Tasso di occupazione (20-64 anni) | 8 |
| Tasso di mancata partecipazione al lavoro | 8 |
| Occupati non regolari | 8 |
| Rapporto tra i tassi di occupazione (25-49 anni) delle donne con figli in eta` prescolare e delle donne senza figli | 5 |
| Divario retributivo di genere | 8 |

Fonte: elaborazioni su dati Italiadomani e Istat, SDGs

---

[17]La documentazione relativa è disponibile sia presso il sito di Italia domani che presso il sito dell'Istat.

[18]In questa prospettiva è utile anche ricordare come la legge di bilancio italiana preveda l'utilizzo di 12 indicatori di Benessere e sostenibilità (Bes) rispetto ai quali misurare gli impatti della finanza pubblica. I 12 indicatori Bes considerati dalla legge di bilancio sono: 1) Reddito disponibile lordo corretto pro capite; 2) Disuguaglianza del reddito netto (s80/s20); 3) Indice di povertà assoluta; 4) Speranza di vita in buona salute alla nascita; 5) Eccesso di peso; 6) Uscita precoce dal sistema di istruzione e formazione; 7) Tasso di mancata partecipazione al lavoro, con relativa scomposizione per genere; 8) Rapporto tra tasso di occupazione delle donne 25-49 anni con figli in età prescolare e delle donne senza figli; 9) Indice di criminalità predatoria; 10) Indice di efficienza della giustizia civile; 11) Emissioni di $CO_2$ e altri gas clima alteranti; 12) Indice di abusivismo edilizio.

Ad esempio, gli indicatori comuni C1 e il C7 appaiono trasversali e non immediatamente riconducibili agli obiettivi propri della componente delle politiche del lavoro. Anche l'indicatore C9 non appare utilizzabile ai fini di un monitoraggio sulle politiche del lavoro a meno di non affiancarlo con una misura dell'occupazione delle imprese che beneficiano di un sostegno. infine, l'indicatore C14 nel suo complesso appare parzialmente riferibile al monitoraggio delle politiche del lavoro, presentando un discreto livello di eterogeneità poiché riferito complessivamente al valore assoluto di giovani che ricevono sostegno, inteso in senso ampio sia per la partecipazione a percorsi educativi, di formazione o di politiche di accompagnamento al lavoro.

Le sovrapposizioni tra i due sottoinsiemi proposti riguardano prevalentemente gli indicatori C10 e C11. L'indicatore C10, Numero di partecipanti in un percorso di istruzione o di formazione è prossimo al tasso di partecipazione alla formazione continua monitorato all'interno del sistema SDGs. Quest'ultima misura, raccordata a livello europeo e basata sull'indagine trimestrale delle forze di lavoro, permette un confronto tra gli andamenti dell'Italia e quelli degli altri paesi europei (Figura 4). In particolare, nel periodo 2014-2021 il tasso di partecipazione italiana si mantiene su livelli significativamente inferiori a quello dell'area euro (3,4 punti percentuali la distanza nel 2019), segnando tuttavia un recupero nell'ultimo anno disponibile (1,5 p.p. la distanza).

L'ancoraggio al sistema statistico europeo permette anche di espandere l'analisi, considerando le ripartizioni geografiche italiane. Nel corso del 2021, il 9,9% degli individui di età compresa fra 25 e 64 anni hanno svolto almeno una attività formativa nelle 4 settimane precedenti l'intervista a sintesi di andamenti fortemente differenziati sul territorio, con valori marcatamente inferiori nel mezzogiorno, anche se in significativo recupero nel 2021 (Figura 5).

La questione che si pone è se questi valori possano essere in qualche modo comparabili a quelli dell'indicatore C10. Una prima considerazione riguarda la popolazione di riferimento che per il tasso di partecipazione riguarda la fascia di età 25-64, mentre per l'indicatore C10 sembra riferito al totale della popolazione. Parallelamente, esiste una differenza nel denominatore utilizzato che riguarda nel primo caso la popolazione nella fascia 25-64 (circa 31,7 milioni di residenti al 2019) e, nel secondo, l'intera popolazione (59,8 milioni di residenti). Per fornire una misura della differenza al momento esistente tra i due indicatori è possibile considerare i valori assoluti. Nel 2021 il valore del tasso di partecipazione (9,9%) esprime circa 3,1 milioni di persone, un valore significativamente distante dalle 165mila persone come numero associato all'indicatore C10 per il primo semestre 2022.

L'esempio può essere esteso alle altre componenti, offrendo un quadro articolato delle possibili relazioni tra gli indicatori comuni e quelli SDGs selezionati dalla mappatura rilasciata da Istat e RGS. Ad eccezione dell'indicatore C9, riferito alle imprese beneficiarie, è possibile ipotizzare una prima matrice di relazioni in grado di

Figura 4: Tasso di partecipazione alla formazione continua (ultime 4 settimane) per Area euro e Italia. Anni 2014-2021

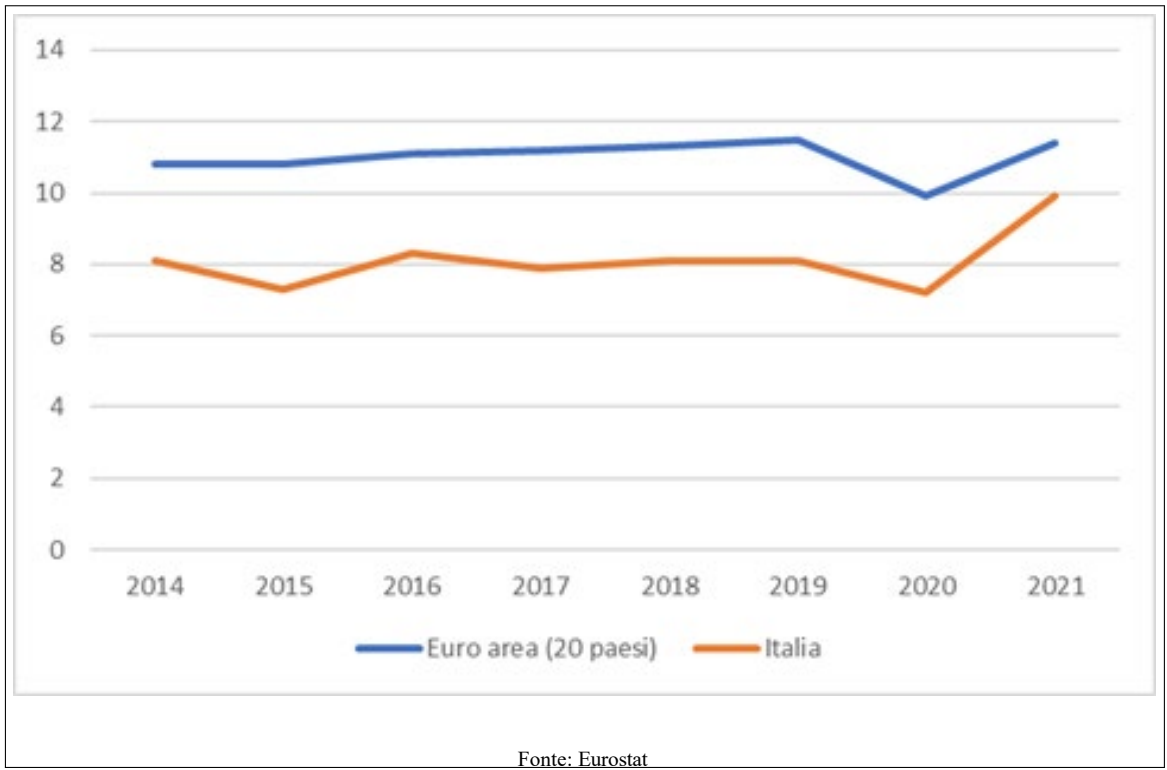

Fonte: Eurostat

Figura 5: Tasso di partecipazione alla formazione continua (ultime 4 settimane) per ripartizione geografica. Anni 2018-2021 (valori percentuali)

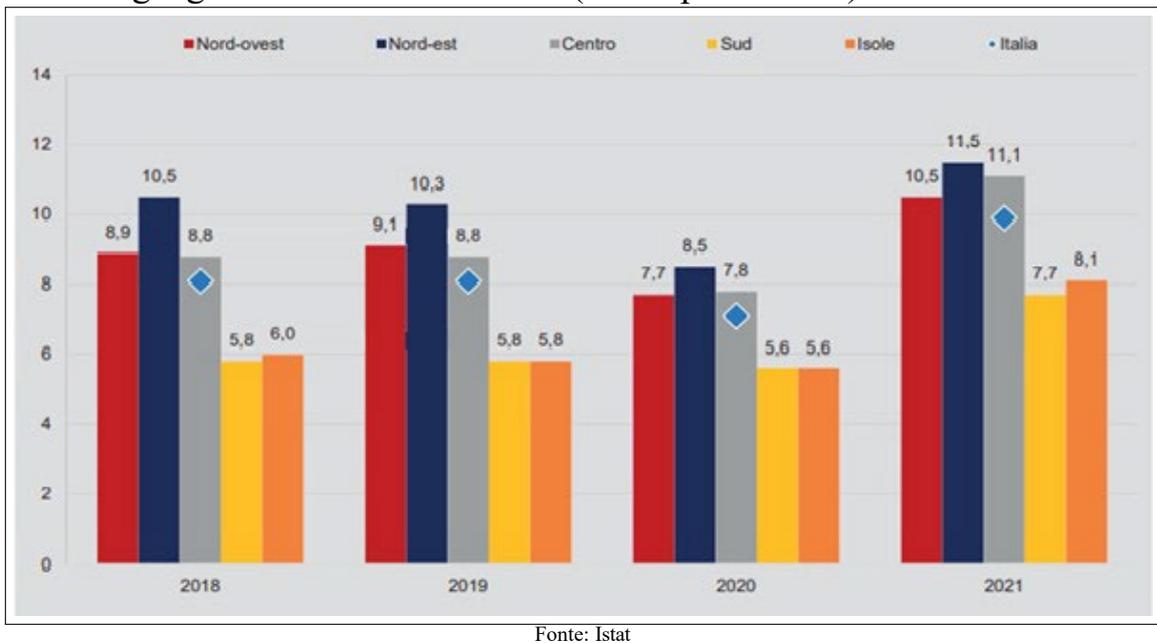

Fonte: Istat

associare a ciascuno dei 14 indicatori comuni un indicatore SDGs di riferimento e il relativo goal (Tabella 8).

Tabella 8: Associazione tra 14 indicatori comuni e 13 indicatori SDGs

| Ind. comuni | *framework* SDGs | |
|---|---|---|
| Codice | Descrizione indicatore | Goal |
| C1 | Intensità energetica | 7 |
| C2 | Energia elettrica da fonti rinnovabili | 7 |
| C3 | Quota di autovetture elettriche o ibride di nuova immatricolazione | 7 |
| C4 | Popolazione esposta al rischio di frane | 13 |
| C5 | Famiglie con connessione a banda larga fissa e/o mobile | 9 |
| C6 | Imprese con attività innovative di prodotto e/o processo | 9 |
| C7 | Persone che hanno interagito online con la Pubblica Amministrazione o con i gestori dei servizi pubblici gestori dei servizi pubblici | 17 |
| C8 | Intensità di ricerca | 9 |
| C9 | – | – |
| C10 | Partecipazione alla formazione continua | 4 |
| C11 | Tasso di mancata partecipazione al lavoro | 8 |
| C12 | Posti letto in degenza ordinaria in istituti di cura pubblici e privati | 3 |
| C13 | Posti autorizzati nei servizi socio educativi (asili nido e servizi integrativi per la prima infanzia) | 4 |
| C14 | Quota della spesa pubblica per misure occupazionali e per la protezione sociale dei disoccupati rispetto alla spesa pubblica | 8 |

Fonte: elaborazioni su dati Italiadomani e Istat, SDGs

Questa prima ipotesi di raccordo richiede ulteriori approfondimenti, ma segnala potenzialità di *fine-tuning* e di integrazione tra indicatori di monitoraggio utilizzati dalle amministrazioni italiane, maggiormente orientato ad una lettura più articolata dei risultati raggiunti dagli RRF.

## 3.3 La mappatura tra PNRR e gli indicatori SDGs

Nel settembre 2022, l'Istat, in collaborazione con la Ragioneria generale dello Stato (RGS), ha diffuso la prima versione della mappatura tra ciascuna sub-misura nell'articolazione del PNRR (missioni/componenti/misure/sub-misure) e uno o più indicatori del *framework* di Benessere (Bes) e Sviluppo sostenibile (SDGs). Più in dettaglio, questo approccio ha permesso di identificare 64 indicatori Bes-SDGs, molti dei quali attribuibili a più di una delle 285 sub-misure (Tabella 9)[19].

La mappatura realizzata, oltre ad evidenziare il numero limitato di indicatori disponibili per il monitoraggio delle missioni M3 (mobilità sostenibile) e M6 (salute), permette di collegare il PNRR agli obiettivi (goal) dell'Agenda 2030, assegnando alle sub-misure i goal dei relativi indicatori SDGs (Tabella 10)[20].

---

[19]Il relativo cruscotto è disponibile a questo link.

[20]Si rimanda alla nota 8 per una descrizione dettagliata dei 17 goal.

Tabella 9: Mappatura del PNRR con gli indicatori Bes ed SDGs

|    | Missioni | Importo (Mld) | | Misure | Sub-misure | Indicatori |
|----|----------|---------------|---|--------|------------|------------|
| M1 | Digitalizzazione | 21,0% | 40,29 | 48 | 108 | 16 |
| M2 | Transizione ecologica | 31,0% | 59,46 | 56 | 64 | 23 |
| M3 | Mobilità sostenibile | 13,3% | 25,40 | 21 | 30 | 7 |
| M4 | Istruzione e ricerca | 16,1% | 30,88 | 35 | 35 | 23 |
| M5 | Inclusione e coesione | 10,4% | 19,85 | 21 | 31 | 29 |
| M6 | Salute | 8,2% | 15,63 | 10 | 17 | 8 |
|    |          |       | 191,50 | 191 | 285 | (*) 64 |

Il valore totale esprime il numero complessivo di indicatori SDGs coinvolti mentre per ogni missione sono riportate le singole occorrenze, che possono essere ripetute.
Fonte: cruscotto Istat-RGS

Le occorrenze espresse dalla mappatura in termini di goal risultano in linea con i pilastri che caratterizzano il PNRR. In particolare, gli indicatori afferenti al Goal 9, *Costruire una infrastruttura resiliente e promuovere l'innovazione e una industrializzazione equa, responsabile e sostenibile*, sono presenti in tutte le missioni sottolineando la rilevanza della dimensione della resilienza e della sostenibilità come evidenziato dal Pilastro 3 *Crescita intelligente, sostenibile e inclusiva*. Anche gli indicatori riferiti al Goal 7, *Assicurare a tutti l'accesso a sistemi di energia economici, affidabili, sostenibili e moderni*, appaiono ampiamente diffusi tra le missioni, sottolineando l'importanza del Pilastro 1, *Transizione verde*.

Tra gli altri goal, la mappatura evidenzia la diffusione tra le missioni degli indicatori dei Goal 4, *Istruzione di qualità per tutti*, Goal 11, *Rendere le città e gli insediamenti umani inclusivi, sicuri, resilienti e sostenibili* e Goal 13 *Adottare misure urgenti per combattere il cambiamento climatico e le sue conseguenze*.

Tabella 10: Mappatura delle missioni del PNRR con i goal SDGs

|    | Missioni | | | | | | | | | | | | | Goal SDGs |
|----|----------|---|---|---|---|---|---|---|---|---|---|---|---|---|
| M1 | Digitalizzazione      |   |   | 4 |   |   | 7 | 8 | 9 | 10 | 11 | 12 | 13 |    | 16 |
| M2 | Transizione ecologica | 1 |   |   |   | 6 | 7 |   | 9 |    | 11 | 12 | 13 | 14 |    |
| M3 | Mobilità sostenibile  |   |   |   |   |   |   |   | 9 |    | 11 |    | 13 |    |    |
| M4 | Istruzione e ricerca  |   | 3 | 4 | 5 |   | 7 | 8 | 9 |    |    |    |    |    |    |
| M5 | Inclusione e coesione | 1 | 3 | 4 | 5 |   | 7 | 8 | 9 | 10 | 11 |    | 13 |    | 16 |
| M6 | Salute                | 1 | 3 | 4 |   |   | 7 |   | 9 |    |    |    |    |    |    |

Fonte: cruscotto Istat-RGS

La Tabella 9 può essere letta anche in senso inverso (SDGs → PNRR), tramite il concetto di associazione *prevalente*. Per ogni sub-misura viene specificato l'indicatore statistico di riferimento tra quelli individuati realizzando un legame tra sub-misura e goal. Nel caso in cui nessun indicatore sia associato alla sub-misura viene individuato almeno un goal di riferimento. Questo permette di attribuire gli importi delle sub-misure ai relativi goal, ottenendo l'articolazione finanziaria descritta nella Tabella 11.

Il Goal 9 *Costruire una infrastruttura resiliente ecc.* e il Goal 7 *Assicurare a tutti l'accesso a sistemi di energia economici ecc.* rappresentano le dimensioni più rilevanti della sostenibilità in termini delle risorse del PNRR.

Tabella 11: Articolazione finanziaria del PNRR per goal SDGs

|    | Goal                                                  | Importo (Mil) |        |     |
|----|-------------------------------------------------------|---------------|--------|-----|
| 1  | Porre fine ad ogni forma di povertà                   | 2.650         | 1,4%   |     |
| 2  | Porre fine alla fame                                  | 0             | 0,0%   | (-) |
| 3  | Assicurare la salute e il benessere                   | 13.506        | 7,1%   |     |
| 4  | Fornire un'educazione di qualità                      | 16.738        | 8,7%   |     |
| 5  | Raggiungere l'uguaglianza di genere                   | 85            | 0,0%   |     |
| 6  | Garantire la gestione sostenibile dell'acqua          | 2.380         | 1,2%   |     |
| 7  | Assicurare l'accesso a sistemi di energia economici   | 34.314        | 17,9%  |     |
| 8  | Incentivare una crescita economica sostenibile        | 8.648         | 4,5%   |     |
| 9  | Costruire una infrastruttura resiliente               | 57.444        | 30,0%  |     |
| 10 | Ridurre le disuguaglianze fra le Nazioni              | 800           | 0,4%   |     |
| 11 | Rendere le città inclusive e sostenibili              | 20.726        | 10,8%  |     |
| 12 | Garantire modelli sostenibili di produzione           | 3.925         | 2,0%   |     |
| 13 | Adottare misure per il cambiamento climatico          | 19.711        | 10,3%  |     |
| 14 | Utilizzare le risorse marine in modo sostenibile      | 400           | 0,2%   |     |
| 15 | Favorire un uso sostenibile dell'ecosistema terrestre | 0             | 0,0%   | (-) |
| 16 | Promuovere società pacifiche e inclusive              | 10.173        | 5,3%   |     |
| 17 | Rafforzare l'attuazione dello sviluppo sostenibile    | 0             | 0,0%   | (-) |
|    |                                                       | 191,50        | 100,0% | (*) |

(-) Ad alcuni goal non viene associato direttamente un importo.

(*) La somma dei singoli importi per goal potrebbe differire dal valore totale a causa dell'arrotondamento considerato.

L'importanza dell'individuazione dei 64 indicatori SDGs risiede anche nella opportunità offerta dall'utilizzo dei dati in serie storica per il periodo 2010-2022 per le tre principali dimensioni dei *framework* Bes/SDGs: territorio, genere ed età (Figura 6). Il territorio rappresenta la dimensione più popolata per gli indicatori, seguita da genere ed età. Questa distribuzione rispecchia l'orientamento statistico nei rapporti Bes ed SDGs di fornire quasi sempre gli incroci genere/territorio e genere/età e di mostrare almeno il dettaglio territoriale per ogni indicatore.

La profondità e la coerenza del sistema informativo Bes-SDGs si è riflessa nell'articolazione del cruscotto Istat-RGS che permette la visualizzazione dei dati degli indicatori. La Figura 7 mostra l'infografica della dimensione Territorio per l'indicatore *Famiglie con connessione a banda larga fissa e/o mobile* (SDG-99) collegato alla sub-misura *Sanità connessa* (M1C2I3.01.04). Ogni elemento grafico descrive un valore presente nelle tavole statistiche che può essere visualizzato dinamicamente con il cursore.

Figura 6: Distribuzione degli indicatori per territorio, sesso ed età

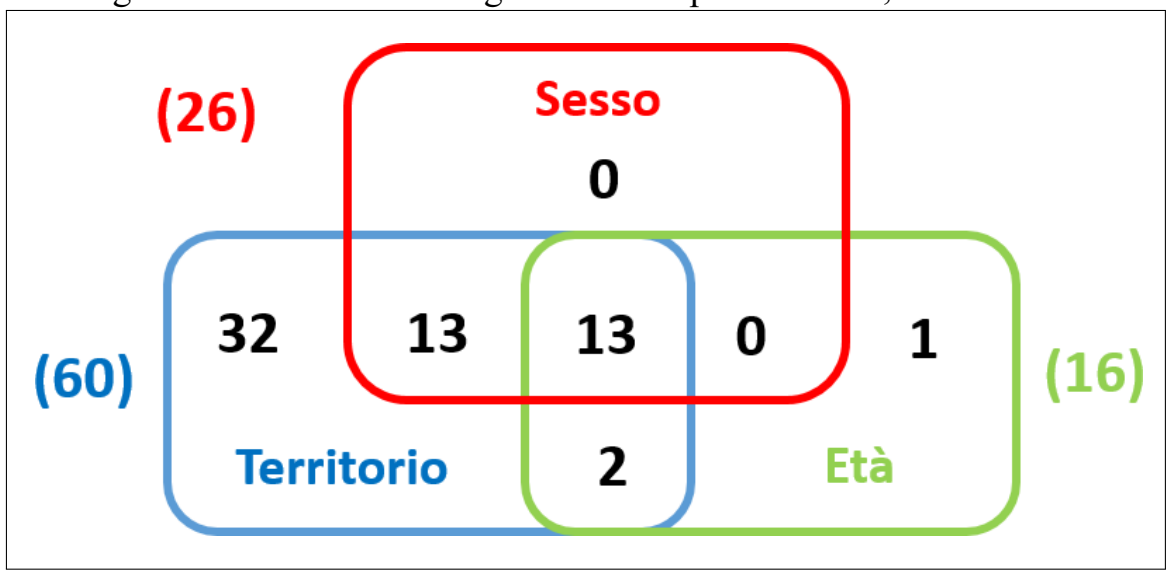

Fonte: elaborazioni su sistema informativo Istat Bes-SDGs

Figura 7: Famiglie con connessione a banda larga fissa e/o mobile per territorio associato alla sub-misura M1C2I3.01.04

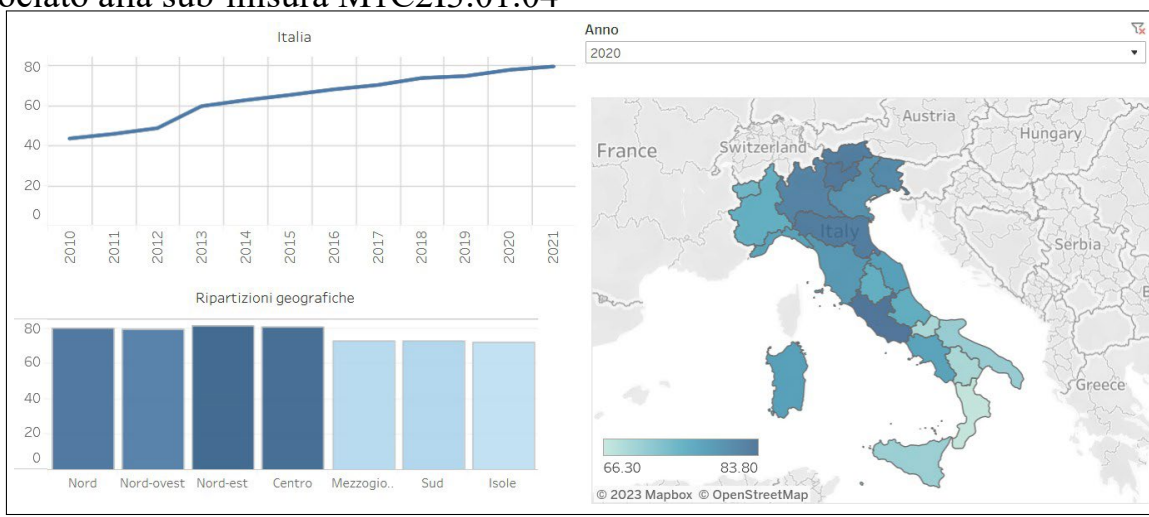

Fonte: cruscotto Istat-RGS

L'infografica relativa alle altre dimensioni è più semplificata: la Figura 8 riporta la dimensione età per la sub-misura *Housing First (innanzitutto la casa) e stazioni di posta* (M5C2I1.03.00) e l'indicatore *Percentuale di persone che vivono in abitazioni con problemi strutturali o problemi di umidità* (SDG-225), mentre la Figura 9 illustra la dimensione genere per la sub-misura *Innovazione e tecnologia della microelettronica* (M1C2I2.01.00) e l'indicatore *Ricercatori (in equivalente tempo pieno)* (SDG-84).

Figura 8: Percentuale di persone che vivono in abitazioni con problemi strutturali o problemi di umidità per età associato alla sub-misura M5C2I1.3

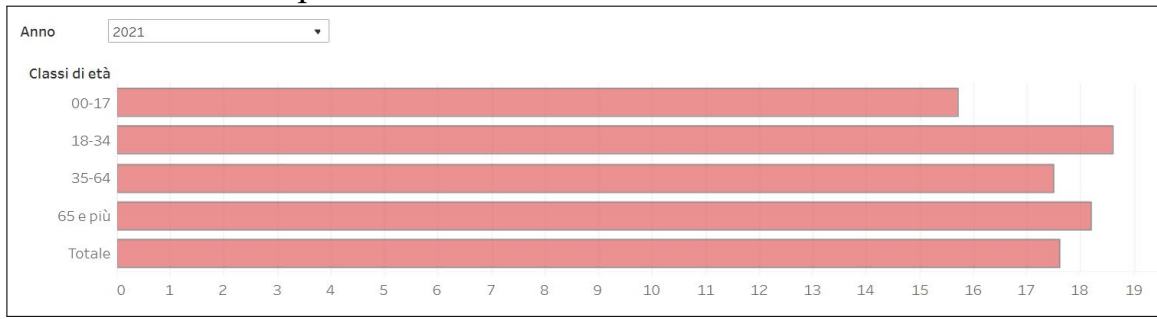

Fonte: cruscotto Istat-RGS

Figura 9: Infografica Ricercatori (in equivalente tempo pieno) per genere - associato con sub-misura M1C2I2.01

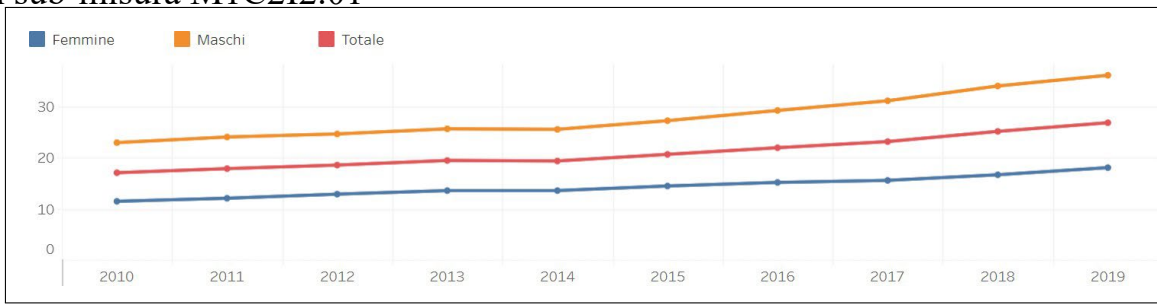

Fonte: cruscotto Istat-RGS

## 4 L'indice composto SDGs-RRF

Il raccordo tra i 14 indicatori comuni e i 13 indicatori SDGs, definiti per ciascun paese UE, così come presentato in Tabella 8, permette di ampliare l'analisi sui risultati dei piani nazionali, legandola al *framework* dello sviluppo sostenibile. In particolare, ad un piano nazionale può essere associata la serie storica di un indice composto degli indicatori SDGs, denominato SDGs-RRF, la cui stima permette un'analisi degli andamenti temporali nazionali anche rispetto alla convergenza tra paesi.

### 4.1 Il quadro metodologico

È indubbio che approcci di sintesi basati su una *dashboard* di indicatori (come, ad esempio, per i già citati rapporti annuali Istat sugli SDGs, [Istat, 2022], e sul Bes, [Istat, 2023]), benché adatti a rappresentare fenomeni multidimensionali, non rispondono però alle pressanti richieste di semplificazione da parte degli stakeholders. In effetti, i concetti complessi sono più facilmente comunicabili nella forma di un unico valore, piuttosto che attraverso un insieme ampio di indicatori (Greco et al.

2019, Saltelli 2007, Stiglitz et al. 2018). Non a caso, negli ultimi due decenni l'utilizzo di indici compositi ha registrato una ampia diffusione (Becker et al. 2017, Greco et al. 2019). Il premio Nobel A.K. Sen, inizialmente scettico circa la costruzione di indici compositi, nell'analizzare l'impatto mediatico dell'indice di sviluppo umano delle Nazioni Unite (*HDI – Human Development Index*, UNDP 2016) ha successivamente notato l'elevato influsso dell'HDI nel promuovere il dibattito sullo sviluppo umano (Sharpe 2004). In generale, però, non vi sono opinioni unanimi né sull'appropriatezza degli indici compositi, né sulla migliore metodologia da adottare per la loro costruzione [21].

Vi sono diverse strade percorribili nella costruzione di un indice composito che richiedono alcune scelte rispetto ai metodi di normalizzazione e aggregazione (cfr. anche Decancq and Lugo 2013, OECD and JRC 2008). Inoltre, come dimostrato anche dall'esperienza dell'HDI, il successo di un indice composito non è ascrivibile solo al rigore statistico, ma anche alla sua semplicità e comunicabilità.

Nella nostra proposta, si è scelto di conservare il pieno vantaggio di una misura cardinale; la prima operazione da effettuare riguarda la normalizzazione degli indicatori elementari, per depurarli delle differenti unità di misura e – in parte – delle differenti variabilità, applicando un metodo classico *min-max* tra 0 e 100, fissando il minimo agli "zeri naturali" (*natural zeroes*) e il massimo agli "obiettivi ambiziosi" (*aspirational targets*). Nell'applicazione corrente la definizione degli zeri naturali e degli obiettivi ambiziosi è stata realizzata a partire dalle serie storiche per gli anni disponibili (dal 2000) e per i paesi disponibili (nell'ambito dell'Unione Europea, considerando anche Svizzera, Islanda, Norvegia e Gran Bretagna), calcolando i minimi e i massimi delle distribuzioni, o se più estremi, il primo e il terzo quartile traslati per 1,5 volte la distanza interquartile [22]. In questo modo sono state evitate scelte arbitrarie [23].

La procedura di normalizzazione *min-max* tra 0 e 100 presenta alcuni vantaggi: è coerente con il calcolo dell'HDI; il *range* scelto non penalizza la movimentazione dell'indice composito; l'utilizzo dei *goalposts* evita il ricalcolo delle normalizzazioni di anno in anno; non si deve scegliere un anno di riferimento già in fase di normalizzazione, rendendo più trasparente l'interpretazione dell'indice composito ed evitando effetti indesiderati sulla funzione di aggregazione (Bacchini et al. 2020a); gli indicatori polarizzati negativamente (in cui cioè una diminuzione dei valori corrisponde a un aumento del benessere) possono essere facilmente invertiti, senza le distorsioni introdotte dalle trasformazioni dei numeri indice.

---

[21] Per un sunto del dibattito in atto sulla bontà degli indici compositi si veda Bacchini et al. 2021, par. 3; per una rassegna dei principali metodi di normalizzazione ed aggregazione, nonché esempi di pratiche già sperimentate anche in ambito internazionale si veda Bacchini et al. 2020a.

[22] È uso considerate un valore non anomalo se entro il primo quartile meno 1,5 volte la distanza interquartile e il terzo quartile più 1,5 volte la distanza interquartile.

[23] Per quanto riguarda l'HDI, i due *goalposts* sono scelti specificamente per ciascuno degli indicatori elementari, attraverso considerazioni basate su studi scientifici.

Lo strumento utilizzato per la sintesi degli indicatori normalizzati è la media geometrica, in coerenza con l'HDI, rispettosa delle proprietà desiderabili in un indice composito (continuità, monotonia stretta, cfr. Mauro et al. 2018, Casadio Tarabusi and Guarini 2013). In aggiunta, la media geometrica non è completamente compensativa, così da tener conto degli squilibri tra le categorie di indicatori su cui si regge l'indice composito (Bacchini et al. 2020a, Casadio Tarabusi and Guarini 2013, UNDP 2010). Infatti, in un indice composito, ogni indicatore è introdotto per rappresentare un aspetto rilevante di un fenomeno, perciò una perfetta sostituibilità tra fattori potrebbe non essere auspicabile. In particolare, la media geometrica presenta gradi minori di sostituibilità per valori più piccoli e sbilanciati: questo vuol dire che i paesi che vogliono vedere aumentare il loro *ranking* hanno un maggior incentivo ad affrontare in maniera decisa i problemi (riflessi negli indicatori con punteggi più bassi, Mauro et al. 2018). Nello stesso tempo, la media geometrica assicura che un determinato incremento o decremento percentuale in ciascuno dei singoli indicatori (una volta normalizzati) abbia esattamente lo stesso impatto sull'indice composito.

L'indice si può utilizzare per confronti sia spaziali che temporali. In particolare, si noti che, utilizzando come funzione di aggregazione la media geometrica, la variazione temporale dell'indice composito si può scomporre nelle variazioni temporali degli indicatori normalizzati. Anche il rapporto dei valori di due paesi diversi è scomponibile nei rapporti degli indicatori normalizzati dei due paesi. In questo modo si possono meglio comprendere le determinanti delle differenze e delle variazioni dell'indice composito.

## 4.2  Il calcolo dell'indice SDGs-RRF

Applicando agli indicatori SDGs selezionati i metodi di normalizzazione e aggregazione descritti nel paragrafo precedente, si ottiene un indice composito di facile interpretazione: il valore dell'indice in ogni anno e per ogni paese è compreso tra 0 e 100, e rappresenta per quell'anno e per quel paese la percentuale di raggiungimento complessivo degli obiettivi ambiziosi.

Per quanto riguarda il calcolo effettivo dell'indice composito, la Tabella 12 riporta, per l'insieme dei paesi europei specificati, gli indicatori SDGs considerati con le unità di misura, l'anno di aggiornamento, alcune statistiche (Q1 e Q3 rappresentano il primo e il terzo quartile), gli "zeri naturali" e gli "obiettivi ambiziosi" (identificati rispettivamente da G1 e G2)[24]. Rispetto alla prima mappatura effettuata (Tabella 8),

---

[24]L'aggregazione dei dati al 2021, ha comportato un'ipotesi di invarianza rispetto all'ultimo dato disponibile riferito

riferita solo alle evidenze disponibili per l'Italia, alcuni degli indicatori individuati non risultano disponibili per qualche paese: in alcuni casi sono state scelte delle *proxy*, mentre nel caso degli indicatori comuni C4 e C6, non è stato possibile trovare una *proxy* appropriata e omogenea per i paesi considerati.

Tabella 12: Indicatori SDGs selezionati per gli indicatori comuni (IC): descrizione, statistiche e goalposts

| IC | Descrizione SDGs | Unità di misura | Agg. | Min | Q1 | Mediana | Q3 | Max | G1 | G2 |
|---|---|---|---|---|---|---|---|---|---|---|
| C1 | Intensità di energia (-) | KGOE/1000€ | 2021 | 40,9 | 118,7 | 163,5 | 248,9 | 769,6 | 0,0 | 769,6 |
| C2 | Quota di energia elettrica da fonti rinnovabili | percentuale | 2021 | 0,1 | 8,6 | 16,9 | 27,9 | 85,8 | 0,0 | 85,8 |
| C3 | Quota dei veicoli a emissioni zero tra le autovetture di nuova immatricolazione | percentuale | 2020 | 0,0 | 0,0 | 0,0 | 0,5 | 51,6 | 0,0 | 51,6 |
| C5 | Famiglie con connessione a banda larga | percentuale | 2021 | 3,8 | 40,7 | 69,3 | 85,9 | 98,6 | 0,0 | 100,0 |
| C7 | Internet: interazione con amministrazioni pubbliche | percentuale | 2021 | 4,9 | 28,1 | 48,1 | 62,9 | 94,2 | 0,0 | 100,0 |
| C8 | Spesa interna lorda per ricerca e sviluppo su PIL | percentuale | 2021 | 0,2 | 0,8 | 1,3 | 2,1 | 3,9 | 0,0 | 4,2 |
| C10 | Partecipazione alla formazione continua | percentuale | 2022 | 0,9 | 5,0 | 8,1 | 17,1 | 37,3 | 0,0 | 37,3 |
| C11 | Tasso di disoccupazione (-) | percentuale | 2022 | 2,0 | 6,1 | 8,0 | 10,6 | 27,8 | 0,0 | 27,8 |
| C12 | Posti letto in ospedale | su 100.000 ab. | 2020 | 205,4 | 364,9 | 536,0 | 667,9 | 911,6 | 0,0 | 1122,5 |
| C13 | Bambini di età inf. a 3 anni in servizi socio-educativi | percentuale | 2022 | 0,0 | 11,8 | 25,7 | 39,3 | 78,0 | 0,0 | 80,6 |
| C14 | Spesa per protezione sociale su spesa pubblica | percentuale | 2021 | 13,9 | 31,5 | 35,9 | 40,0 | 45,8 | 13,9 | 52,7 |

Legenda: il segno meno per l'indicatore comune C1 e C11 indicano che a un aumento del valore corrisponde un peggioramento (polarità negativa); in tutti gli altri casi la relazione è diretta
Fonte: Eurostat

Nella Figura 10 si rappresenta l'andamento dell'indice composito SDGs-RRF per i quattro principali paesi dell'area Euro (Germania, Spagna, Francia, Italia) per il periodo 2014-2021[25].

Anche tenendo conto dei valori degli indicatori normalizzati e dell'indice composito SDGs-RRF per i quattro paesi, all'inizio e alla fine del periodo considerato (Tabella 13), l'Italia e la Spagna partono da una situazione di forte ritardo rispetto a Germania e Francia: l'indice composito SDGs-RRF per Italia e Spagna è inferiore di circa un terzo rispetto alla Germania, e di circa il 40% rispetto alla Francia (in termini assoluti l'indice composito per l'Italia e la Spagna si trova 11-12 punti sotto la Germania e 15-16 punti sotto la Francia). In particolare, in Italia, rispetto alla Germania, è significativamente debole la quota dei veicoli a emissione zero (per l'indicatore comune C3), l'interazione con le amministrazioni pubbliche tramite

---

al 2020 per gli indicatori sulla quota dei veicoli a emissioni zero tra le autovetture di nuova immatricolazione (afferente a C3) e dei posti letto in ospedale (afferente a C12).

[25]I goalpost utilizzati per il calcolo sono quelli elaborati per l'insieme dei paesi europei specificati (Tabella 12).

Figura 10: Indice composito SDGs-RRF per Francia, Germania, Italia e Spagna. Anni 2014-2021

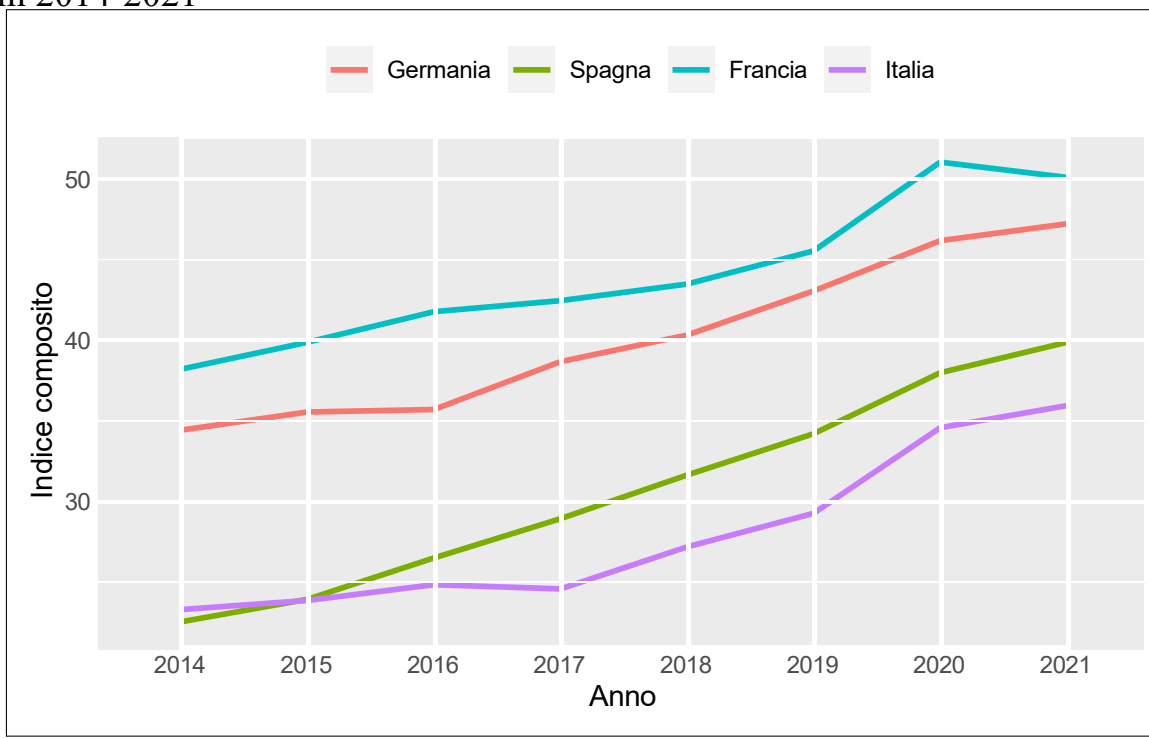

Fonte: elaborazioni su dati Eurostat

internet (C7), la spesa per R&D (C8) e il numero dei posti letto in ospedale (C12). Nel confronto con la Francia, l'Italia mostra ulteriori debolezze anche per quanto riguarda la partecipazione alla formazione continua (C10) e i bambini negli asili nido (C13).

In Spagna l'indice SDGs-RRF evidenzia un progressivo miglioramento, sin dal 2015, e, nel periodo considerato, cresce del 76,5%, superando l'Italia e riducendo il divario sia con la Germania che con la Francia. Nel caso italiano, invece, l'indice rimane stabile tra il 2014 e il 2017, per poi aumentare fino al 2021, ma con una decelerazione tra il 2020 e il 2021 (peraltro comune anche agli altri paesi). Tra il 2014 e il 2021 l'indice SDGs-RRF per l'Italia sale del 54% (quasi 13 punti in termini assoluti), soprattutto in virtù di un aumento percentuale particolarmente consistente della quota dei veicoli a emissioni zero (C3), ma anche dell'utilizzo di Internet nell'interazione con le amministrazioni pubbliche (C7) e dei bambini negli asili nido (C13). Quasi tutti gli indicatori elementari crescono nel periodo, ad eccezione dell'intensità di energia (C1) che rimane stabile e del numero dei posti letto in ospedale (C12), che scende dell'1,2%.

Nel 2021, l'indice composito è compreso tra i 36,1 punti dell'Italia e i 50,3 punti della Francia (un divario di 14,2 punti, in diminuzione rispetto a quello del 2014,

Tabella 13: Indicatori SDGs normalizzati e indice composito SDGs-PNRR. Francia, Germania, Italia e Spagna. Anni 2014 e 2021

| Indicatore comune | Indicatore SDGs normalizzato | | | | | | | |
|---|---|---|---|---|---|---|---|---|
| | Francia | | Germania | | Italia | | Spagna | |
| | 2014 | 2021 | 2014 | 2021 | 2014 | 2021 | 2014 | 2021 |
| C1 | 83,9 | 85,8 | 84,9 | 86,9 | 87,2 | 87,2 | 84,1 | 85,3 |
| C2 | 16,7 | 22,5 | 16,8 | 22,3 | 19,9 | 22,2 | 18,5 | 24,2 |
| C3 | 1,2 | 13,0 | 0,6 | 12,4 | 0,2 | 4,5 | 0,2 | 4,1 |
| C5 | 76,7 | 88,1 | 86,6 | 88,8 | 71,1 | 88,5 | 73,0 | 95,9 |
| C7 | 63,8 | 80,7 | 52,7 | 50,3 | 23,0 | 33,9 | 49,0 | 68,7 |
| C8 | 52,8 | 52,3 | 68,2 | 74,1 | 31,7 | 35,0 | 29,3 | 33,8 |
| C10 | 49,3 | 29,5 | 21,4 | 20,6 | 21,7 | 26,5 | 27,1 | 38,6 |
| C11 | 62,9 | 71,6 | 83,1 | 86,7 | 53,6 | 65,8 | 11,9 | 46,8 |
| C12 | 55,2 | 51,2 | 73,3 | 69,6 | 28,6 | 28,3 | 26,4 | 26,3 |
| C13 | 49,0 | 70,9 | 34,1 | 39,0 | 28,4 | 41,5 | 45,8 | 68,7 |
| C14 | 74,5 | 72,2 | 73,5 | 69,4 | 71,2 | 73,0 | 66,3 | 69,1 |
| Indice composito | 38,4 | 50,3 | 34,6 | 47,4 | 23,4 | 36,1 | 22,7 | 40,0 |

Fonte: elaborazione su dati Eurostat

pari a 15,7 punti tra Francia e Spagna). La riduzione del divario relativo tra Italia e Francia è dovuta, in particolare, all'avvicinamento dei livelli della partecipazione alla formazione continua (C10), non solo per un miglioramento italiano, ma anche per un vistoso decremento francese nel periodo pandemico (2020-2021).

Il processo di convergenza tra i principali quattro paesi dell'area Euro appare piuttosto limitato sul periodo, usando il *set* di indicatori SDGs individuati.

# 5 Considerazioni conclusive

L'entità dei fondi messi a disposizione dei piani nazionali RRF richiede uno sforzo rivolto al monitoraggio e alla valutazione del loro impatto nella desiderata direzione del rafforzamento di un sistema economico, sociale e ambientale improntato alla resilienza.

L'attuale sistema predisposto dalla Commissione europea, denominato *Scoreboard*, include sia la definizione di *milestone* e *target* che un insieme di 14 indicatori comuni. Nel presente lavoro si suggerisce un percorso rivolto a un'integrazione più articolata tra questo sistema degli indicatori comuni (esplicitamente legato ai piani nazionali dell'RRF) e il *framework* SDGs. La proposta qui presentata offre una ulteriore opportunità in termini di valutazione, permettendo di sfruttare le dimensioni temporali, territoriali e di genere, ampiamente disponibili nel sistema SDGs.

Il lavoro sviluppa una metodologia di raccordo tra i due framework, che viene illustrata dapprima per la sua adattabilità alle missioni e componenti del PNRR (fornendo un esempio per una singola missione/componente) e successivamente fornisce una mappatura tra i 14 indicatori comuni ed un sottoinsieme di indicatori SDGs come loro *proxy*.

Utilizzando questi indicatori, per i quali sono presenti dati per il periodo 2014- 2021, è stato costruito un originale indice composito, denominato SDGs–RRF per Francia, Germania, Italia e Spagna. Gli indici compositi nazionali elaborati mostrano una modesta riduzione delle distanze tra i paesi nel periodo considerato; al tempo stesso, il quadro della loro evoluzione permette di rappresentare un *benchmark* di riferimento per i successivi valori degli indici stessi che rifletteranno l'impatto degli RRF nazionali sugli indicatori SDGs e, quindi, la rispondenza agli obiettivi generali stabiliti dal Dispositivo per la Ripresa e Resilienza. In quest'ottica, l'esperienza italiana è di particolare valore, in considerazione dell'inserimento degli indicatori di benessere e sostenibilità all'interno dei documenti di finanza pubblica e della disponibilità del cruscotto PNRR-SDGs predisposto da Istat e Ragioneria Generale dello Stato.

Ulteriori applicazioni e l'approfondimento metodologico dell'approccio illustrato, come, ad esempio, la sua robustezza a diversi criteri di composizione dell'indice, potrebbero fornire elementi utili al monitoraggio dell'impatto dei piani nazionali, favorendo i confronti con le analisi contenute nei *Country Report* elaborati dalla Commissione europea.

Più in generale, questo lavoro vuole contribuire all'ampliamento della riflessione necessaria per la costruzione di un sistema di valutazione armonizzato sugli effetti dei piani RRF nazionali, di cui il sistema dei 14 indicatori comuni è solo un tassello.

# Riferimenti bibliografici